\documentclass[twocolumn,showpacs,amsmath,nofootinbib]{revtex4-2}

\usepackage{float} 
\usepackage{amsmath}
\usepackage{amsfonts}
\usepackage{amssymb}
\usepackage{amsthm}
\usepackage{graphicx}
\usepackage[colorlinks,citecolor=blue]{hyperref}
\usepackage{color}
\usepackage{flafter}
\usepackage[normalem]{ulem}

\usepackage[toc,page]{appendix}
\usepackage{makecell}
\usepackage{slashed}
\usepackage{tikz-cd}
\usepackage{makecell}
\usepackage{subcaption}
\usepackage{graphicx}
\usepackage{xr}
\usepackage[font=small,labelfont=bf,
justification=justified,
format=plain]{caption}

\allowdisplaybreaks

\definecolor{OliveGreen}{rgb}{0,0.6,0}
\definecolor{Orange}{rgb}{1.00, 0.65, 0}
\definecolor{Grey}{rgb}{0.43, 0.5, 0.5}

\newcommand{\Fig}[1]{Fig.~\ref{#1}}
\newcommand{\Eq}[1]{Eq.(\ref{#1})}
\newcommand{\nn}{\nonumber\\}

\newcommand{\be}{\begin{eqnarray}}
\newcommand{\ee}{\end{eqnarray}}
\newcommand{\bpm}{\begin{pmatrix}}
\newcommand{\epm}{\end{pmatrix}}

\newcommand{\Avg}[1]{\langle #1 \rangle} 
\newcommand{\comment}[1]{}

\begin{document}
\title{Theory of an infinitely anisotropic phase of a two-dimensional superconductor}
\author{Zi-Xiang Li}
\affiliation{Beijing National Laboratory for Condensed Matter Physics and Institute of Physics,
Chinese Academy of Sciences, Beijing 100190, China\\
University of Chinese Academy of Sciences, Beijing 100049, China.}
\author{Steven A. Kivelson}
\affiliation{Department of Physics, Stanford University, Stanford, California 94305, USA.}
\author{Dung-Hai Lee}
\affiliation{Department of Physics, University of California, Berkeley, CA 94720, USA.\\
Materials Sciences Division, Lawrence Berkeley National Laboratory, Berkeley, CA 94720, USA.}

\begin{abstract}
We propose a simple model 
of a two-dimensional ``locally smectic superconductor'' that exhibits power-law non-linear I-V characteristics
 with different powers for current applied in two orthogonal directions. We discuss the potential relevance of this model to recent experimental findings on interfacial superconductivity 
in EuO/KTaO$_3$(110) (Nat. Phys. {\bf 20}, 957 (2024)), where the apparent critical temperature ($T_c$) 
has been shown experimentally to depend on the current direction.
\end{abstract}
\maketitle

\section{Introduction} 
Over the past two decades, interfacial superconductivity has garnered significant interest, with notable examples including superconductivity observed 
in SrTiO$_3$/LaAlO$_3$
 \cite{Reyren2007}, 
 FeSe/SrTiO$_3$ 
 \cite{Xue2012}, 
 and EuO/ KTaO$_3$\cite{Liu2021} interfaces. Remarkably, in Ref.\cite{XH2024}, it is reported that the superconducting state at the EuO/KTaO$_3$ (110) interface displays different apparent critical temperatures (T$_c$) depending on the direction of the probing current (with the current magnitude in the range of $10^{-3}-10^{-2}$ of the ``critical current''). 
In this paper,
motivated by these experiments,
we propose a theoretical model  that exhibits a novel finite temperature ``infinitely anisotropic smectic superconducting'' phase, the conjectured existence of which  can account for the observed phenomena.\\

KTaO$_3$ is a cubic perovskite insulator, and EuO is a ferromagnetic insulator with a Curie temperature 
of approximately 73 K. The (110) interfacial system hosts a two-dimensional electron gas (2DEG) whose density can be tuned by varying the growth parameters. It is effectively orthorhombic with a readily detectable resistance anisotropy at room temperature, and principle axes $[1,\bar 1 ,0]$ and $[0,0,1]$ (which we will refer to as the $x$ and $y$ axes, respectively).   Below the Curie temperature of EuO, the magnetic moments lie in the (110) plane.\footnote{It is unclear to us what orientation of the moments is preferred within this plane.} 
Interfacial superconductivity arises for a range of areal electron density ($n_s$) of the 2DEG.  Surprisingly,   when $n_s$ is lower than approximately $8\times 10^{13}$ cm$^{-2}$, the apparent superconducting resistive transition temperature 
 depends on the direction of the probing current.
 For instance, for $n_s \approx 6.8\times 10^{13}$ cm$^{-2}$, T$_c \approx 450$ mK when the current is passed in the
 $x$ direction, whereas T$_c \approx 250$ mK for current in the 
 $y$ direction. In contrast, 
 when $n_s$ exceeds $8\times 10^{13}$ cm$^{-2}$, T$_c$ is the same regardless of the direction of the probing current.

Since the superconductivity in question is genuinely two-dimensional, the transition temperature (T$_c$) should be governed by the Kosterlitz-Thouless (KT) theory, which describes the unbinding of superconducting vortex-antivortex (V-AV) pairs. 
 Below the KT transition, an applied current exerts a Magnus force that pulls apart 
 previously bound V-AV pairs, leading to a voltage drop in the direction of the current\cite{halperin,review}.  This results in a non-linear voltage-current characteristic, $ V\sim  (I/I_c)^\alpha$,  where $I_c$ is 
 a characteristic ``critical current,'' and the exponent $\alpha$ is temperature dependent. In the conventional KT theory, $\alpha=3$ 
 just below the KT temperature and diverges as $T\rightarrow 0$. 
The apparent dependence of T$_c$ on the current direction suggests that the ease with which the applied current pulls the V-AV pairs apart varies with the current direction. 
Experimentally, for a fixed $I/I_c (<<1)$, 
 there exists a temperature at which the differential resistance, $ R\sim (I/I_c)^{\alpha-1}$,  drops below the minimum measurable value. 
To the extent that this temperature is not too strongly $I$ dependent, it would naturally be identified as the superconducting $T_c$.  If the exponent 
$\alpha$
differs for two orthogonal current directions, this apparent resistance-vanishing temperature can be different. This is the central idea 
that motivates the present study.\\

Naively, one might think that an anisotropic (i.e. nematic) superconductor  
 with an anisotropic phase stiffness tensor, $\rho_{ab}$, 
could explain the observed behavior. However,
 the strength of the logarithmic attraction between vortices and anti-vortices is determined by the geometric mean of 
 $\rho_{aa}$ in the two principle directions, i.e. by $\bar \rho =\sqrt{{\rm Det}[\rho]}$, 
 which implies that the exponent 
$\alpha$ in the I-V characteristic 
would be the same in both orthogonal directions.  A superconducting state with a direction dependent exponent, $\alpha$, is a distinct  phase of matter, characterized by an infinitely anisotropic differential resistivity tensor, $R_{aa}$, in the $I \to 0$ limit. {For example, if $R_{xx}\propto I^{\alpha_1},  R_{yy}\propto I^{\alpha_2}$ with $\alpha_1>\alpha_2$ then $R_{xx}/R_{yy} \to 0$ as $I\to 0$.}  In this paper, we define 
an anisotropic classical XY model with first harmonic couplings in one direction and second harmonic in the other, show that it exhibits the requisite infinitely anisotropic KT phase, and finally describe the sort of microscopic considerations that could generate such an effective model.

\section{The model}

The (classical) model we shall consider
is
\be
H=-
J_1\sum_{i,j}\cos(\theta_{i,j}-\theta_{i+1,j}) -J_2\cos 2(\theta_{i,j}-\theta_{i,j+1}).\nn
\label{model}\ee
Here i,j are the x and y coordinates of a square lattice site, $\theta$ is a phase (representing the local  superconducting phase), and $J_1,J_2>0$ are the microscopic phase stiffness constants. 
This Hamiltonian supports 
both  integer and half-integer vortices. In Fig.1 we show pair consisting of a half-integer/integer vortex and anti-vortex (V-AV).  
In the half-integer case, there is a line defect connecting the pair across which the phase jumps by $\pi$;  this  leads to a linear confining potential unless the pair are separated purely in the $x$ direction.  There is, of course, still a logarithmically growing confining potential between vortex pairs, but while this is the dominant interaction between integer pairs independent of the direction of their relative displacement, it is only the case for half-integer pairs separated (primarily) in the $x$ direction.
\begin{figure}[H]
  \centering
		\includegraphics[scale=0.28]{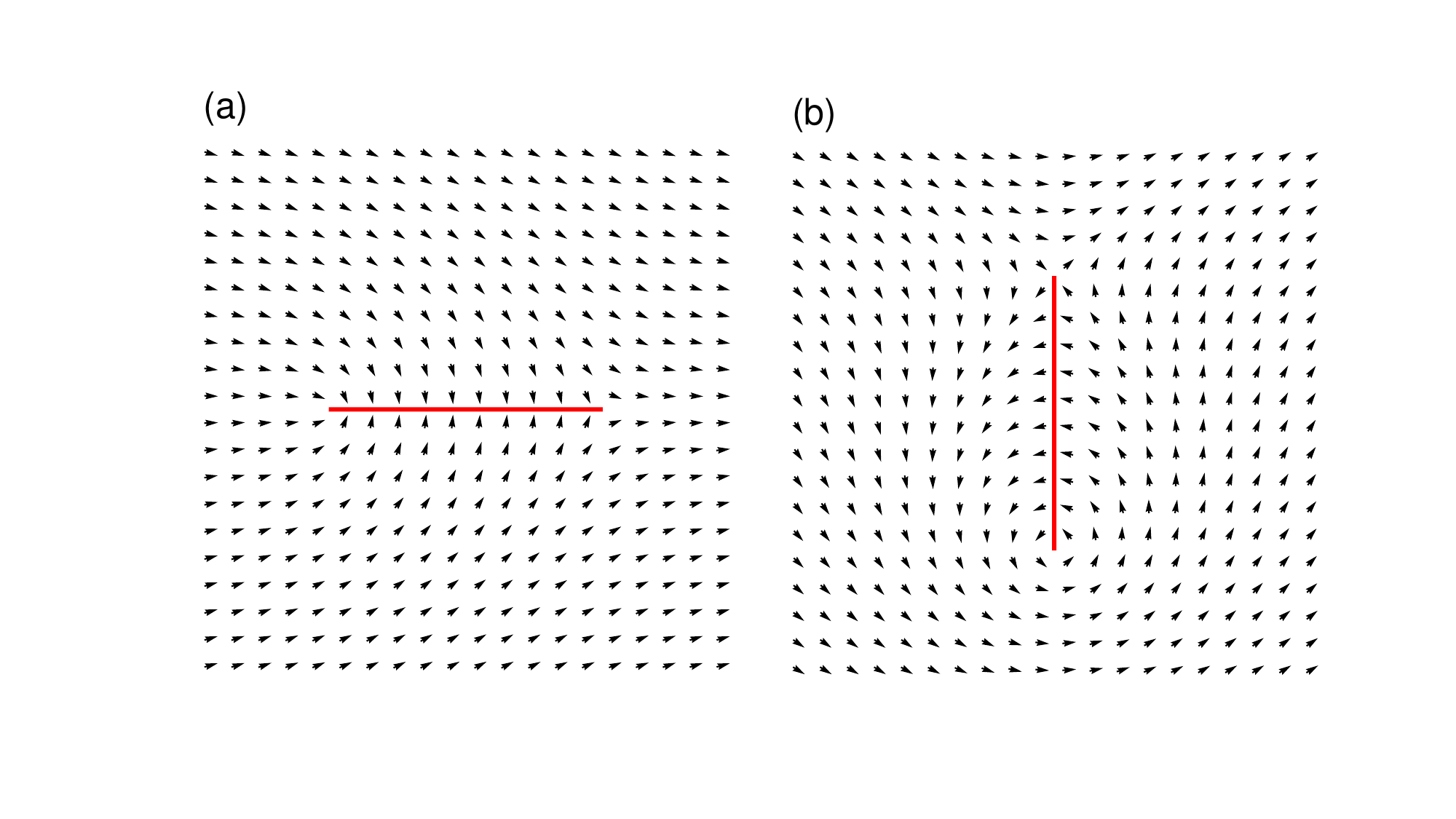}
		\caption{(a) A half-integer vortex anti-vortex pair separated in the x direction. (b) An integer vortex anti-vortex pair separated in the x direction. The arrows defines the phase angle and the red line segment highlight the separation between the vortices.}
		\label{Fig.1}
	\end{figure}

\section{The I-V characteristics}
	
We now consider  
 the energy of an integer and half-integer V-AV pair, $U_{1/2}$ and $U_{1}$, in the presence of a probing current $I$ 
 as a function of their separation
$d$ in the direction perpendicular to the  current flow.  As long as $d$ is large compared to the vortex core radius, the current exerts an equal and opposite Magnus force on the vortex and the anti-vortex, meaning that there is an induced contribution to the energy proportional to $I$ and $ d$.  For current in the $y$ direction, the result can be expressed as
\be
&&U_{1/2}(d) ={2\pi\bar\rho \over 4}\left[\ln{d\over a}-{I\over I_{1/2}
 }{d\over a}+\epsilon_{1/2}\right]
 \nn&&U_{1}\left({d }
 \right)=2\pi\bar \rho
 \left[\ln{d\over a}-{I\over I_{1,y}}{d\over a}+\epsilon_1\right].
\label{energy}\ee
while for (small) $I$ in the $x$ direction, $U_{1}$ is of the same form but with $I_{1,y} \to I_{1,x}$, but  $U_{1/2}$ is a linearly increasing function of $d$.
Here, $a$ is a characteristic short-distance scale (roughly the vortex core radius), $\epsilon_\alpha$ are unimportant constants, 
and the characteristic current scales,  $I_{1/2}$, $I_{1,y}$ and $I_{1,x}$. 
 The important relative factor $1/4$ in the prefactor of $U_{1/2}$ and $U_1$ is simply the square of the vorticity.
 
Given \Eq{energy}, for any non-zero $I$, there is always an energy gain for making sufficiently far-separated V-AV pairs.  The rate at which this happens is determined by the energy barrier to pair creation, which is determined by the energy maxima, which for $I$ along the $y$ direction occur at $d^*_{1/2} = a ({I_{1/2}/I})$ and $d^*_1= a ({I_{1,y}/I})$, while for $I$ in the $x$ direction, where only integer pair formation is allowed, it occurs at $d^*_1= a ({I_{1,x}/I})$.
The corresponding barrier heights, { $\Delta_{1/2}=U_{1/2}\left(d^*_{1/2}\right)-U_{1/2}\Big(a\Big)$ and $\Delta_{1,y}=U_{1}\left(d^*_{1}
\right)-U_{1}\Big(a\Big)$}  for $I$ in the $y$ direction are 
\be
&&\Delta_{1/2}={2\pi\bar\rho
 \over 4}\left[\ln\left({I_{1/2}
 \over I}\right)-1+\left(\frac{I}{I_{1/2}}\right)\right]\nn
&&\Delta_{1,y}=2\pi\bar \rho
\left[\ln\left({I_{1,y}\over I}\right)-1+\left(\frac {I}{I_{1,y}}\right)\right]
\ee
while for current in the $x$ direction, $\Delta_{1,x}$ is given by the same expression as  $\Delta_{1,y}$ with $I_{1,y} \to I_{1,x}$.  
(In the small $I$ limit, the logarithmic  terms dominate these expressions.)
Note that for  $I$ sufficiently small, $\Delta_{1,y} >\Delta_{1/2}$, meaning that for $I$ along $y$, the dominant process is always generation of half integer V-AV pairs.

At temperature $T$ the thermal disassociation rate is proportional to $e^{-\Delta_{1/2}/k_B T}$ for half-integer vortices and
 $e^{-\Delta_{1}/k_B T}$ for integer vortices. On the other hand the recombination rates of disassociated vortices are proportional to
 $n_{1/2,+}n_{1/2,-}$ and $n_{1,+}n_{1,-}$. As a result in the steady state where the disassociation and recombination rate are equal we have for current along the $y$ and $x$ directions, respectively,
 \be 
 &&n_{1/2,+}=n_{1/2,-}\propto e^{-{\Delta_{1/2}\over 2k_BT}}\propto \left({I\over I_{
 1/2}}\right)^{\pi\bar \rho
 \over 4k_BT}\nn&&{n_{1,+}=n_{1,-}\propto e^{-{\Delta_{1,x}\over 2k_BT}}\propto \left({I\over I_{1,x}}\right)^{\pi\bar \rho
 /k_BT}}\ .\label{density}
 \ee 
Thus the voltage drop  and the associated differential resistance are given by
\be
&&{{V_y(T)\over V_{y,0}}=\left({I\over I_{1/2}}\right)^{1+{\pi\bar\rho(T)\over 4k_BT}};{V_x(T)\over V_{x,0}}=\left({I\over I_{1,x}}\right)^{1+{\pi\bar\rho(T)\over k_BT}}}\nn
&&{{R_{yy}(T)\over R_{y,0}}=\left({I\over I_{1/2}}\right)^{{\pi\bar\rho(T)\over 4k_BT}};{R_{xx}(T)\over R_{x,0}}= \left({I\over I_{1,x}}\right)^{{\pi\bar\rho(T)\over k_BT}}} 
\label{Exy}\ee 
where $V_{x,0},V_{y,0}$ are constants with 
dimensions of voltage, and $R_{x,0},R_{y,0}$ are constants with 
dimensions of resistance. Their respective values depends on microscopic details. The extra power of current in the {voltage}  formula is due to the fact that the induced vortex current is proportional to the applied current. A specific prediction of \Eq{Exy} is that at the  thermodynamic (Kosterlitz-Thouless) transition temperature,  
the V-I scaling is $V_y(T_{\rm KT})\propto I^3$ while $V_x(T_{\rm KT})\propto I^9$. These scaling laws can presumably be checked experimentally.
The anisotropic I-V characteristics in \Eq{Exy} is the main result of this paper.

 From them,  the existence of a direction dependent critical temperature  follows.  In typical experiments, measurements are done at a fixed  (small) current density, $I$, and with a fixed sensitivity in the voltage measurement.  There is thus a minimum resistance, {$R_{min}$}, that becomes the operational definition of zero resistance.  Thus the apparent resistive $T_c$ is  the temperature at which 
\be
R_{aa}(T_c) = R_0 \ (I/I_c)^{\alpha(T_c) -1}= R_{min}
\label{mes2}
\ee
where for $I$ in the $x$ direction, $\alpha(T)-1 =\pi\bar\rho/k_BT$, $I_c=I_{1x}$ and $R_0 = R_{x,0}$, while for $I$ in the $y$ direction, $\alpha(T)-1 =(1/4) \pi\bar\rho/k_BT$, $I_c=I_{1/2}$ and $R_0 = R_{y,0}$.
Defining $T_a$ to be the apparent critical temperature defined in this way, it is clear that $T_y > T_x$, and that the difference $T_y-T_x$ grows with decreasing {$R_{min}$}.

\section{Response to small magnetic field}
At temperatures below the zero-field transition temperature, $T_{KT}$, a small concentration of far separated vortices are induced  with vorticity density equal to $B/\phi_0$, is induced by application of a small transverse magnetic field, $B$.  ($\phi_0=hc/2e$ is the superconducting quantum of flux.)  A state with purely half-integer vortices would thus have a vortex density $n_{1/2} = 2(B/\phi_0)$ and an energy density  ${\cal E} \approx n_{1/2} \frac{2\pi \bar \rho}{4}\left| \ln\left[\sqrt{n_{1/2}a^2}\right]\right|$, while with purely integer vortices the vortex density would be half as large, but the energy density roughly twice larger.  Thus, the magnetic-field induced vortices should be primarily half-integer vortices - increasingly so as $B\to 0$.

In the absence of disorder, at low enough temperature, the field induced vortices likely form an Abrikosov lattice with quasi-long-range order.  However, there is expected to be a  range of temperatures between the melting temperature of the Abrikosov lattice and $T_{KT}$ in which a vortex fluid occurs.  In the presence of disorder, no Abrikosov lattice arises, so the vortices remain mobile (possibly with thermally activated mobility) at all non-zero temperatures.  So long as any field-induced vortices remain mobile, the resulting state has a finite linear-response resistance.  Specifically, vortex motion in the $x$ direction produces a voltage drop in the $y$ direction and vice-versa. However,  half-integer vortices on their own are constrained to move only in the $x$ direction - they can move in the $y$ direction only when two half-integer vortices are close to one another and move together.  Thus, in a state with dominantly half-integer vortices, the linear-response resistance in the $y$ direction, $R_{yy}$, is expected to be proportional to $n_{1/2}$, while $R_{xx} \propto n_{1/2}^2$.  As a result, the resistance anisotropy, $R_{xx}/R_{yy} \to 0$ as $B \to 0$. This is another sense in which the resulting state is expected to be infinitely anisotropic.

\section{The Monte-Carlo results} 

\begin{figure}[H]
    \centering
    \includegraphics[scale=0.4]{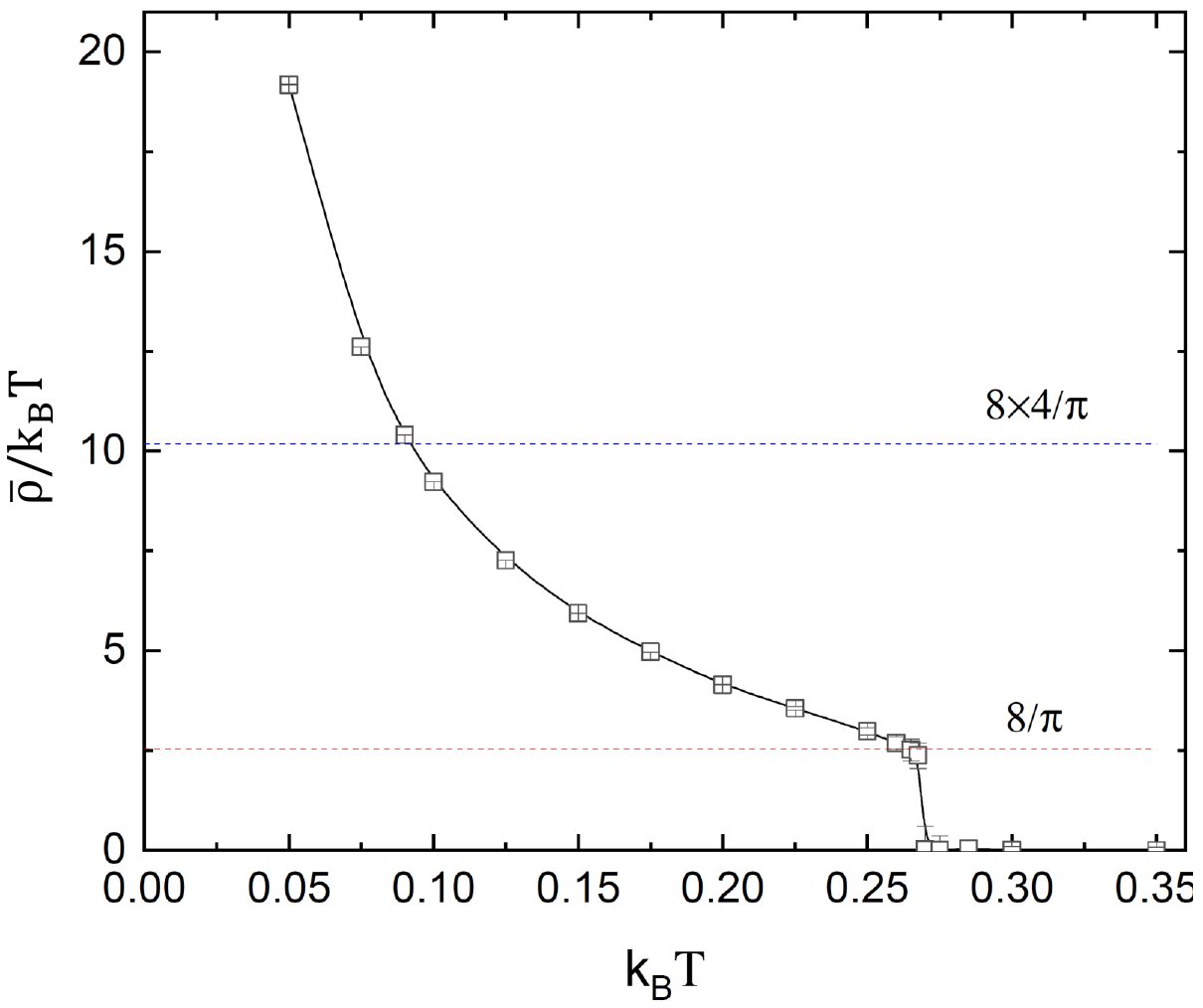}
    \caption{The Monte-Carlo results for the phase stiffness $\rho_{xy}(T)/k_BT$ at thermodynamic limit versus temperature $T$. The red dashed line indicates jump of $\bar\rho$ at $\bar\rho/k_BT = \frac{8}{\pi}$. The blue dashed line is $\bar\rho/k_BT = \frac{4\times8}{\pi}$. }  
    \label{Fig3}
\end{figure}

To compute the phase stiffness $\bar \rho$ as a function of temperature, we performed large-scale Monte-Carlo simulations on the model of \Eq{model}. The microscopic phase stiffness constants in \Eq{model} are fixed as $J_1 = 1.0$ and $J_2 = 0.25$. Here and thereafter, all energies are expressed in units of $J_1$. In  the simulation, we adapted a combination of Wolff's global update and a local update to minimize the statistical errors. The simulation was done on 
 $L\times L$ square lattices with the largest system size $L= 260$, at which
point there are no discernible changes in the results upon further increasing 
$L$. The phase stiffness in the x and y- directions were evaluated 
from:
\be
\rho_{xx} = \frac{J_1}{L^2}\Avg{\sum_{ij} K_x(i,j)} - \frac{J_1^2}{k_BTL^2} \Avg{\big[\sum_{ij} J_x(i,j)\big]^2} \nn
\rho_{yy} = \frac{4J_2}{L^2}\Avg{\sum_{ij} K_y(i,j)} - \frac{4J_2^2}{k_BTL^2} \Avg{\big[\sum_{ij} J_y(i,j)\big]^2}
\ee
where $\Avg{\cdots}$ represents thermal ensemble average, 
$T$ is temperature, $K_x(i,j) = \cos(\theta_{i,j}-\theta_{i+1,j}) $, $K_y(i,j) = \cos(\theta_{i,j}-\theta_{i,j+1}) $ and $J_x(i,j) = \sin(\theta_{i,j}-\theta_{i+1,j}) $, $J_y(i,j) = \sin(\theta_{i,j}-\theta_{i,j+1}) $ . 
To extrapolate $\bar \rho
 (L)$ to the thermodynamic limit, we did a third-order polynomial fitting of $\bar \rho
 (L)$ versus $1/L$, using the results with system sizes ranging from $L=60$ to $L=260$. In \Fig{Fig3} we plot the extrapolated $\bar \rho
 (L\rightarrow\infty)/k_BT$ versus $k_BT$. The results show a sharp drop  $\frac{8k_BT}{\pi}$ at temperature $k_BT_{\rm KT} \approx 0.2675$. According to the KT theory, 
 the $\frac{8k_BT_{\rm KT}}{\pi}$ drop in  $\bar \rho
 $ is 
 the expected result for the unbinding of half-integer V-AV pairs, and  $k_BT_{\rm KT}$ marks the associated  Kosterlitz-Thouless phase transition. 
Note that, there is a {\it single} thermodynamic phase transition temperature $T_{\rm KT}$.  

\begin{figure}[H]
    \centering
    \includegraphics[scale=0.4]{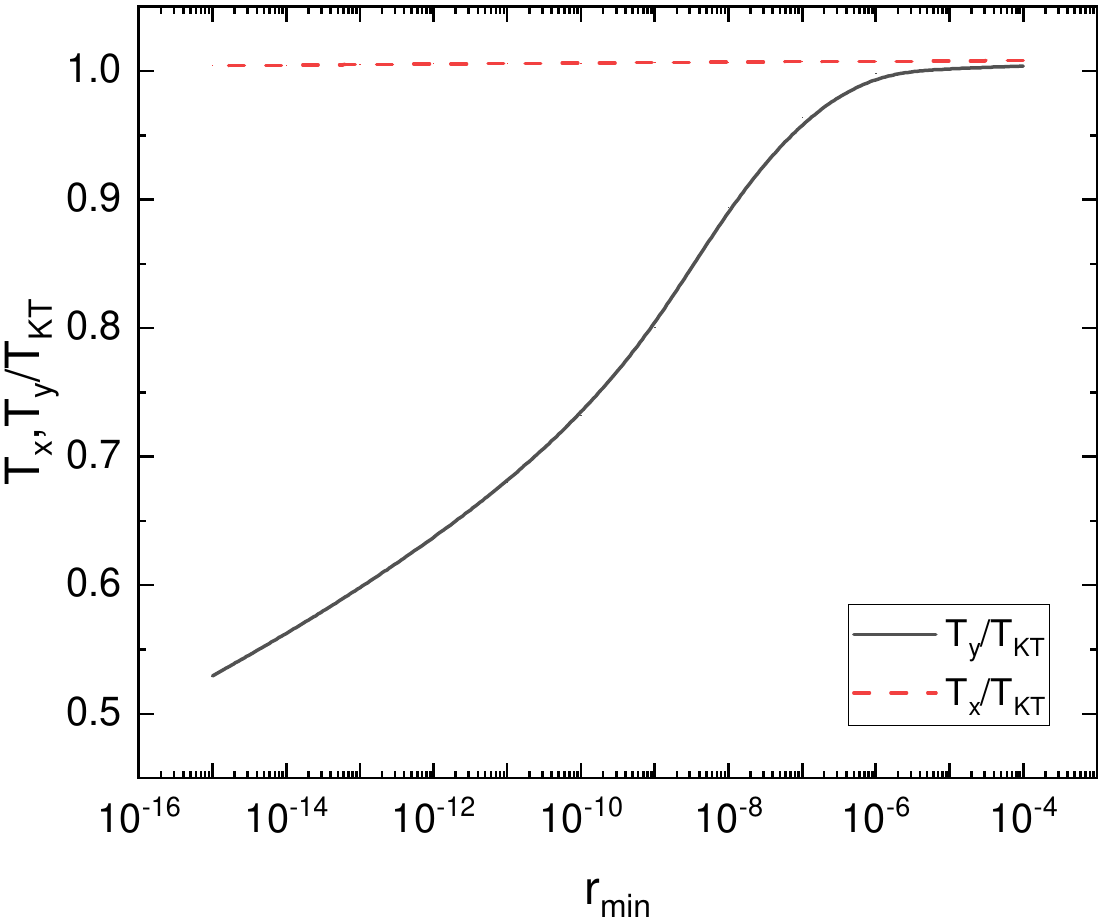}
    \caption{    $T_x$ and $T_y$ (defined in Eq. \ref{mes2})
     as a function of  $r_{\rm min}\equiv R_{min}/R_0$ with $\bar\rho(T)$ from the Monte-Carlo calculations.  The magnitude of the probing current, $I/I_c$, is set to $10^{-3}$ and for simplicity we have taken $I_{1/2}=I_{1x}\equiv I_c$ and $R_{y,0}= R_{1,x}\equiv R_0$.   As discussed in the text, $T_x$, shown as a dashed line, is actually a lower bound on the actual $T_x$.}  
    \label{Fig4}
\end{figure}

However, as shown in Fig. \ref{Fig4}, using the Monte-Carlo results as input, we obtain two distinct ``apparent'' resistive transition temperatures, as defined in \Eq{mes2}. A subtlety is that, because the large power law exponent in $R_{xx}(T)$, 
  for the small value we have taken of $I/I_c=10^{-3}$, at $T_{\rm KT}$, where $R_{xx}/R_0= (I/I_c)^8$, 
  the differential resistance computed in this way is smaller than the lowest value 
of $r_{min}$ shown  in the figure,
i.e.  \Eq{mes2} has no solution for $T_x$ below $T_{\rm KT}$. 
Since the linear-response $R_{xx}$ (i.e. $\lim_{I\to 0} R_{xx}$) vanishes as  $T\to T_{\rm KT}$ from above, 
under these circumstances $T_x$ must satisfy the inequality $T_x\ge T_{\rm KT}$.  
However, without going through further microscopic-dependent analysis 
of $R_{xx}$ above $T_{\rm KT}$,  all we can conclude is that  $T_x \geq T_{\rm KT}$. 
We represent this by
showing $T_x$ as a dashed line in the figure.

\section{Motivation for the model}
Our motivation for considering the model given by \Eq{model} is the following: For strongly correlated systems, such as doped Mott insulators, the competing interaction between the short distance attraction (such as AF exchange) and the long range Coulomb repulsion can results in various kind of electronic liquid crystal phases\cite{KFE1998}.   Among them is a smectic stripe phase which coexists with a superconducting phase. In such phase  an ordered array of one dimensional metallic rivers exhibit Cooper pairing at relatively high temperatures. 

\begin{figure}[H]
  \centering
		\includegraphics[scale=0.27]{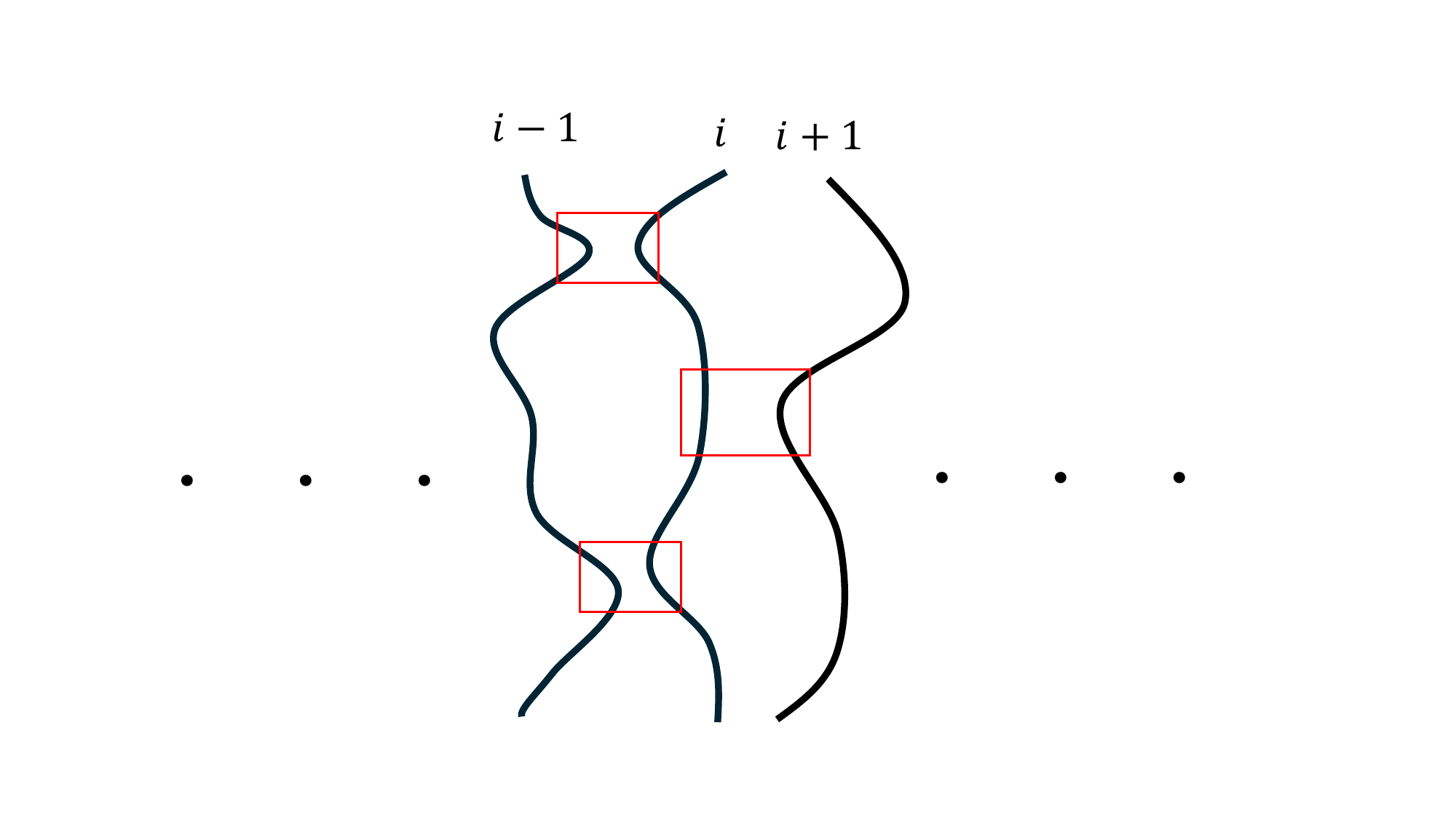}
		\caption{Josephson tunneling (enclosed by the red boxes) between meandering stripes.}
		\label{Fig.2}
	\end{figure}

Motivated by the dependence of the apparent $T_c$ on the current direction, the authors of Ref.\cite{XH2024} conjectured the existence of superconducting stripes.  However, in two dimensions, interactions with quenched disorder precludes the existence of long-range (or even quasi-long range) stripe order, so while we adopt this picture of the local order, at long distances only the orientational (nematic) order of this state (pinned to a specific lattice direction by the orthorhombic symmetry of the interface) is expected to survive.  A cartoon of such a state is shown in Fig. \ref{Fig.2}.

To a picture of self-organized superconducting rivers, we add the observation that the in-plane magnetization makes it plausible that the superconducting state involved resembles a finite momentum Fulde-Ferrell-Larkin-Ovchinnikov (FFLO) like state.  Such a state can be described by local order parameters, $\psi_{n,\pm}(s)$ where $n$ labels the stripe (i.e. it is roughly the $x$ coordinate), $\pm Q$ are the pair-density-wave ordering vector along the stripe, and $s$ is the distance along the stripe. Additional motivation for this picture
 is provided by the works
of Ref.\cite{White2006} and Ref.\cite{Weng2023} where t-J ladders in the presence of Zeeman field
were found to exhibit finite momentum
pairing.

There are several notable features of the inter-stripe coupling of the superconducting order that are implied by this picture.  Firstly, because the Josephson tunneling depends exponentially on distance, it is likely dominated by points of closest approach between neighboring stripes, indicated by the red boxes in Fig. \ref{Fig.2}.  However, because the path-length along two neighboring stripes is different, the lowest order Josephson coupling between the order parameters on neighboring stripes $n$ and $n+1$ is of the form 
\be
-{\cal J}_{\lambda,\lambda^\prime} e^{iQ(\lambda s_n-\lambda^\prime s_{n+1})}\psi_{n,\lambda}^\star(s_n)\psi_{n+1,\lambda^\prime}(s_{n+1}) + {\rm c.c.}
\ee
which then is summed over components, $\lambda \ \& \lambda^\prime=\pm$, and over the points of closest approach with   $s_n$ and $s_{n+1}$ being the coordinate along the respective stripes of the points of close approach.  Here ${\cal J}_{\lambda,\lambda^\prime}$ are positive Josephson couplings (with magnitude that can vary from point to point).  The essential point is that on performing the sum over points of close approach, the average Josephson coupling (with $\psi_{n,Q}(s)$ independent of $s$) vanishes not only for backscattering interactions ($\lambda = -\lambda^\prime$) but also, due to the random distribution of $s_{n}-s_{n+1}$,  for forward scattering interactions as well ($\lambda = \lambda^\prime$).  
This same frustration-induced suppression of inter-stripe coupling applies to any finite $Q$ order along the stripe, and indeed, the same argument was presented in Ref. \cite{KFE1998} as a proposed mechanism for suppression of inter-stripe CDW order.

However, it is possible to make a composite zero momentum order parameter, $\Psi_{n}(s) = \psi_{n,+}(s_n)\psi_{n,-}(s_n)$, for which the coupling between stripes, ${\cal K}$, is of uniform phase.  Manifestly, $\Psi$ is a charge $4e$ order parameter.  Such couplings can occur as a result of coherent interstripe tunneling of pairs-of-pairs\cite{franz,andrew2}, or can be generated in second order perturbation theory, ${\cal K} \sim {\overline{{\cal J}^2}}$, from a model with pure random-in-phase first-order Josephson couplings\cite{spivak,andrew}.  
The essential point is that because it couples composite (charge $4e$) orders on neighboring stripes, in terms of the superconducting phase variables this implies a
$\cos 2(\theta_i-\theta_{i+1})$ coupling rather than the usual $\cos (\theta_i-\theta_{i+1})$. 

One could generalize the present analysis to include coupling of arbitrary high-order zero-momentum order parameters, for example $[\psi_{n,+}(s_n)\psi_{n,-}(s_n)]^2$ (charge $8e$) or $\psi_{n,+}(s_n)\psi_{n,+}^\star(s_n)$ (which creates charge $0e$).  The important point is that all possible zero momentum composite order parameters create an even integer number of Cooper pairs and hence are invariant under $\theta_i \to \theta_i + \pi$.  This is the symmetry that permits the production of half-inteter V-AV pairs along a stripe.

\section{Other examples of infinitely anisotropic phases}
There have been several other cases in which distinct phases of matter with  infinite anisotropy in their response to perturbations has been discussed in various exotic circumstances.  The idea of quantum ($T=0$) ``sliding phases'' in a 2D array of coupled 1DEGs has been explored from several perspectives\cite{sliding1,sliding2,sliding3};  this is closely related to the idea of ``floating phases'' in a layered system consisting of weakly coupled classical XY models\cite{floating1,floating2,floating3}.   However, finding conditions in which these effectively 3D (or 2+1D) examples are absolutely stable to all perturbations at best requires considerable (possibly unrealizable) tuning of parameters.  It is presumably for this reason that no clear experimental evidence exists of such phases in condensed matter systems.  This should be contrasted with the stability of the behavior we have found, so long as some symmetries preclude the generation of any odd-harmonic couplings in one direction.

The one higher dimensional case characterized that is known to be perturbatively stable occurs in a layered system in which each layer is a ``spin-liquid,'' i.e. has an emergent gauge symmetry.  Here the infinite anisotropy of the response functions of such systems is insured by the separate topological quantum numbers associated with each plane.\cite{tfloating1,tfloating2,tfloating3}  As far as experimental realization, it is plagued by the same uncertainties that are associated with all existing searches for spin-liquids in quantum materials.  In both respects, this is a very different case than the present one.  

\section{Conclusion}

While we have analyzed a specific form of an anisotropic XY model, the novel infinitely anisotropic superconducting phase we have unveiled presumably does not depend on details of the model, but only on certain symmetries.  Any model of an XY ferromagnet (or for that matter an XY antiferromagnet on an unfrustrated lattice) that respects those symmetries presumably would support such a phase.  What is essential is that it is an XY model, i.e. invariant under the global transformation 
$\theta_{ij} \to \theta_{ij} + \phi$ for any $\phi$, is even  under a global ``time reversal transformation,'' $\theta_{ij} \to - \theta_{ij}$, and that it has a discrete $\mathbb{Z}_2$ symmetry associated with each fixed column i.e $\theta_{ij} \to \theta_{ij} + \eta_i \pi$ where $\eta_j =0$ or $1$.

We have also reported phenomenlogical reasons to conjecture that such a phase exists for a range of electron-concentrations in the interfacial superconducting phase of EuO/KTaO$_3$.  This conjecture provides a reasonable account of the striking observation of a probe-current-angle dependent value of the apparent resistive transition temperature, $T_c$\cite{XH2024}.  However, whether or not the model we have analyzed is actually realized in that system remains to be seen.  \\

\acknowledgements
D.H.L was  funded by the U. S. Department of Energy, Office of Science, Office of Basic Energy Sciences,
Materials Sciences and Engineering Division under Contract No. DE-AC02-05-CH11231 (Theory of Materials program KC2301). S.A.K was funded, in part, by NSF-BSF award DMR2310312. Z.X.L. was funded by the NSFC under Grant No.~12347107.


\begin{thebibliography}{99}
\bibitem{Reyren2007} N. Reyren {\it et al.} ``Superconducting interfaces between insulating oxides'', Science, {\bf 317},1196-1199 (2007).
\bibitem{Xue2012} Q.-Y. Wang {\it et al} '`Interface-Induced High-Temperature Superconductivity in Single Unit-Cell FeSe Films on SrTiO3'', Chinese Physics Letters {\bf 29}, 037402 (2012).
\bibitem{Liu2021} CJ Liu {\it et al.} ''Two-dimensional superconductivity and anisotropic transport at KTaO$_3$ (111) interfaces'' Science {\bf 371}, 716b(2021).
\bibitem{XH2024} Hua, X., Zeng, Z., Meng, F. et al. ``Superconducting stripes induced by ferromagnetic proximity in an oxide heterostructure,'' Nat. Phys. (2024). https://doi.org/10.1038/s41567-024-02443-x
\bibitem{KFE1998} S. A. Kivelson, E. Fradkin and V. J. Emery ``Electronic liquid-crystal
phases of a doped
Mott insulator'', Nature {\bf 393}, 550 (1998).
\bibitem{White2006} G. Roux {\it et al.} ``Zeeman Effect in Superconducting Two-Leg Ladders: Irrational Magnetization Plateaus
and Exceeding the Pauli Limit'', Physical Review Letters {\bf 97}, 087207 (2006).
\bibitem{Weng2023} H-K Zhang, R-Y Sun, and Z-Y Weng ``Pair density wave characterized by a hidden string order parameter'', Physical Review B {\bf 108}, 115136 (2023).

\bibitem{floating1}  S. E. Korshunov and A. I. Larkin, ``Problem of Josephson-vortex-lattice melting in layered superconductors,''
Phys. Rev. B {\bf 46}, 6395 (1992).

\bibitem{floating2} C. S. O'Hern and T. C. Lubensky, ``Sliding Columnar Phase of DNA-Lipid Complexes,'' Phys. Rev. Lett. {\bf 80}, 4345 (1998).

\bibitem{floating3} C. S. O'Hern, T. C. Lubensky, and J. Toner, ``Sliding Phases in XY Models, Crystals, and Cationic Lipid-DNA Complexes,'' Phys. Rev. Lett. {\bf 83}, 2745 (1999).

\bibitem{sliding1} Emery, V. J. and Fradkin, E. and Kivelson, S. A. and Lubensky, T. C., ``Quantum Theory of the Smectic Metal State in Stripe Phases,'' \prl {\bf 85}, 2160 (2000).
 
\bibitem{sliding2} A. Vishwanath and D. Carpentier, ``Two-Dimensional Anisotropic Non-Fermi-Liquid Phase of Coupled Luttinger Liquids,'' Phys. Rev. Lett. {\bf 86}, 676 (2001).

\bibitem{sliding3} S. L. Sondhi and K. Yang, ``Sliding phases via magnetic fields,'' Phys. Rev. B {\bf 63}, 054430 (2001).

\bibitem{tfloating1} 
P. W. Anderson and Z. Zou, ``"Normal" Tunneling and "Normal" Transport: Diagnostics for the Resonating-Valence-Bond State,''
Phys. Rev. Lett. {\bf 60}, 132 (1988).

\bibitem{tfloating2} Yochai Werman, Shubhayu Chatterjee, Siddhardh C. Morampudi, and Erez Berg, ``Signatures of Fractionalization in Spin Liquids from Interlayer Thermal Transport,'' Phys. Rev. X {\bf 8}, 031064 (2018).

 \bibitem{tfloating3} Trithep Devakul, S. L. Sondhi, S. A. Kivelson, and Erez Berg, ``Floating topological phases,'' 
Phys. Rev. B {\bf 102}, 125136 (2020).

\bibitem{halperin}  Halperin, B.I., and D. R. Nelson, ``Resistive transition in superconducting films,'' J. Low Temp. Phys. {\bf 36}, 599 (1979).

\bibitem{review} For a review, see P. Minnhagen, ``The two-dimensional Coulomb gas, vortex unbinding, and superfluid-superconducting films,'' Rev. Mod. Phys. {\bf 59}, 1001 (1987).

\bibitem{dror}  Dror Orgad, ``Effects of geometrical fluctuations on the transition temperature of disordered quasi-one-dimensional superconductors,''  Phys. Rev. B {\bf 79}, 014509 (2009)

\bibitem{spivak}  A. Zyuzin and B. Spivak, ``Theory of $\pi/2$ superconducting Josephson junctions,'' Phys. Rev. B {\bf 61}, 5902 (2000).

\bibitem{andrew}  Andrew C. Yuan, Yaar Vituri, Erez Berg, Boris Spivak, and Steven A. Kivelson, ``Inhomogeneity-induced time-reversal symmetry breaking in cuprate twist junctions,'' Phys. Rev. B {\bf 108}, L100505 (2023).

\bibitem{franz}  Tarun Tummuru, Stephan Plugge, and Marcel Franz, ``Josephson effects in twisted cuprate bilayers,''
Phys. Rev. B {\bf 105}, 064501 (2022).

\bibitem{andrew2}  
Andrew C. Yuan and Steven A. Kivelson, ``Phase sensitive information from a planar Josephson junction,'' arXiv:2404.11657.
\end{thebibliography}
\end{document}